\documentclass[aps,preprint]{revtex4}
\usepackage[dvips]{graphicx}

\usepackage{color}

\usepackage{amsmath}    
\usepackage{amsfonts}
\usepackage{amssymb}
\usepackage{fancybox}
\begin{document}

\title{Spin coherent states in NMR quadrupolar system: experimental and theoretical applications}

%
%
\author{R. Auccaise} \email{rauccais@cbpf.br}
\affiliation{
Instituto de F\'{i}sica de S\~{a}o Carlos, Universidade de S\~{a}o Paulo,\\
P.O. Box 369, S\~{a}o Carlos 13560-970,  S\~{a}o Paulo, Brazil.}
\author{E. R. deAzevedo} 
\affiliation{
Instituto de F\'{i}sica de S\~{a}o Carlos, Universidade de S\~{a}o Paulo,\\
P.O. Box 369, S\~{a}o Carlos 13560-970,  S\~{a}o Paulo, Brazil.}
\author{E. I. Duzzioni}
\affiliation{
Instituto de F\'{i}sica, Universidade Federal de Uberl\^{a}ndia,\\
P.O. Box 593, Uberl\^{a}ndia 38400-902, MG, Brazil.}
\author{T. J. Bonagamba}
\affiliation{
Instituto de F\'{i}sica de S\~{a}o Carlos, Universidade de S\~{a}o Paulo,\\
P.O. Box 369, S\~{a}o Carlos 13560-970,  S\~{a}o Paulo, Brazil.}
\author{M. H. Y. Moussa} 
\affiliation{
Instituto de F\'{i}sica de S\~{a}o Carlos, Universidade de S\~{a}o Paulo,\\
P.O. Box 369, S\~{a}o Carlos 13560-970,  S\~{a}o Paulo, Brazil.}


\begin{abstract}
Working with nuclear magnetic resonance (NMR) in quadrupolar spin systems,
in this paper we transfer the concept of atomic coherent state to the
nuclear spin context, where it is referred to as pseudo-nuclear spin
coherent state (pseudo-NSCS). Experimentally, we discuss the initialization
of the pseudo-NSCSs and also their quantum control, implemented by polar and
azimuthal rotations. Theoretically, we compute the geometric phases acquired
by an initial pseudo-NSCS on undergoing three distinct cyclic evolutions: $%
i) $ the free evolution of the NMR quadrupolar system and, by analogy with
the evolution of the NMR quadrupolar system, that of $ii)$ single-mode and $%
iii)$ two-mode Bose-Einstein Condensate like system. By means of these
analogies, we derive, through spin angular momentum operators, results
equivalent to those presented in the literature for orbital angular momentum
operators. The pseudo-NSCS description is a starting point to introduce the
spin squeezed state and quantum metrology into nuclear spin systems of
liquid crystal or solid matter.
\end{abstract}


\maketitle

\section{Introduction\label{sec:level1}}

In a seminal work, R. J. Glauber \cite{glauber1963} proposed the concept of
a coherent state as an eigenstate of the annihilation operator of the
harmonic oscillator. This concept was extended to atoms as the well-known
atomic coherent states (ACS) \cite{arecchi1972,perelomov}. These coherent
states, used in many physical processes in the fields of quantum
electrodynamics \cite{raimond2001}, trapped ions \cite{leibfried2003} and
Bose-Einstein condensates (BECs) \cite{leggett2001}, are of fundamental
importance because they minimize the Heisenberg uncertainty \cite
{arecchi1972}. The minimization of uncertainty tends to amplify the quantum
effects on the macroscopic scale. One physical system that manifests quantum
effects at a macroscopic level is the BEC. In this connection, an
interesting application of the ACS was developed by Chen et al. \cite
{chen2004}, who computed geometric phases of a coupled two-mode BEC system
for an adiabatic and cyclic time evolution. Other applications are the
generation of macroscopic quantum superpositions \cite{gordon1999} and
dynamic control \cite{bargill2009,grond2009} in BECs.

The topic of geometric phase has recently received considerable attention
and it has been employed increasingly in quantum information processing \cite
{jones2000,ota2009PRA1,chen2009,niu2010}. The quantum version of geometric
phases was first reported and studied by M.V. Berry in the 1980's \cite
{berry1984} and implemented, thereafter, in many physical systems \cite
{tomita1986,bhandari1988,leek2007}. A general discussion of geometric phases
acquired by pure states was published by N. Mukunda and R. Simon \cite
{mukunda1993}. In this new approach, the authors took advantage of quantum
kinematic concepts, so that this formalism could be applied in a general
context, such as in any smooth open path defined by unit vectors in a
Hilbert space. Many applications using this formalism have appeared, for
instance in noncyclic geometric quantum computation \cite{wang2009} and,
more recently, nonadiabatic geometric quantum gates using composite pulses 
\cite{ota2009PRA1}, both of these applications being discussed in the NMR
context. Other applications of the Mukunda-Simon approach have been the
study of geometric phases in a nonstationary superposition of atomic states
induced by an engineered reservoir \cite{prado2009}, and also the study of
the effect of geometric phases on the pseudospin dynamics of two coupled
BECs \cite{duzzioni2007} and of a spin-orbit-coupled BEC \cite{larson2009}.
Likewise, NMR experiments in systems with two coupled spins and quadrupolar
spin systems have been published, in which the geometric phase of an
adiabatic cyclic evolution of an initial quantum state is analyzed \cite
{suter1987}. Other potential NMR applications of adiabatic \cite
{jones2000,ekert2000} and nonadiabatic \cite{xiang2001,zhu2002A,zhu2002B}
geometric quantum computation have also been reported.

These advances inspired us to study two subjects related to NMR quadrupolar
systems. The first is the possibility of transferring the definition of ACS
to a nuclear spin state, which will be referred to as a pseudo-nuclear spin
coherent state (pseudo-NSCS). The second concerns an application of
pseudo-NSCSs to compute geometric phases in three different configurations: $%
i)$ the free evolution of the NMR quadrupolar system \cite{abragam1994} and,
by analogy of the evolutions of the NMR quadrupolar system, that of $ii)$
single- and $iii)$ two-mode BEC-like system \cite{auccaise2009,milburn1997}.
Under this cyclic evolution, we apply the Mukunda and Simon formalism and
obtained theoretical results analogous to those of Zhang et al. in Ref. \cite
{zhang2006PRL}. Special attention was given to configurations $ii)$ and $%
iii) $, in which we took the advantage of the mapping between the
quasiparticle description and pseudo-spin momentum operators for any spin
value $I$.

In general, the present paper presents experimental results and theoretical
applications of the mapping between the atomic and nuclear spin scenarios.
This will be a basis for future experimental implementations of quantum
control, and also to establish an NMR quadrupolar system as a workbench for
a BEC system.

This paper is organized as follows: In section \ref{sec:TSCSsection}, the
main concepts and a review of ACS are presented, while in section \ref
{sec:NMRQuadrupolar} the NMR quadrupolar system is briefly described.
Section \ref{sec:experimentalprocesso} is devoted to the experimental
procedure to initialize the pseudo-NSCSs and describes the quantum control
of polar and azimuthal rotations. In section \ref{sec:aplicaciones},
theoretical applications of the pseudo-NSCS based on the geometric phase
concept are elaborated for configurations $i)$, $ii)$ and $iii)$ and in
section \ref{sec:discusiones} we report our conclusions.

\section{Atomic coherent states\label{sec:TSCSsection}}

In the theory of ACS, a Bloch vector $\mathbf{n}=\left( \sin \theta \cos
\varphi , \right.$  $\left. \sin \theta \sin \varphi ,\cos \theta \right) $ is transformed by
the action of a rotation operator $\mathbf{R}_{\theta ,\varphi }=e^{-i\theta
\left( \mathbf{J}\cdot \mathbf{m}\right) }$. The operator $\mathbf{R}%
_{\theta ,\varphi }$ represents a rotation of angle $\theta $ around an axis
defined by $\mathbf{m}=\left( \sin \varphi ,-\cos \varphi ,0\right) $, where 
$\mathbf{J\cdot m}=\mathbf{J}_{x}\sin \varphi -\mathbf{J}_{y}\cos \varphi $.{%
\ The operator $\mathbf{J}$ signifies the total angular momentum, with
components $\mathbf{J}_{x}$, $\mathbf{J}_{y}$, and $\mathbf{J}_{z}$, such
that $j$ is the total angular momentum quantum number and its projection
corresponds to $m=-j,-j+1,\ldots ,j-1,j$.} After a simple algebraic
procedure, the application\ of $\mathbf{R}_{\theta ,\varphi }$ to the
fundamental state denoted by $\left| j,-j\right\rangle $ produces an excited
state represented by \cite{perelomov,inomata,arecchi1972}: 
\begin{eqnarray}
\left| \zeta \left( \theta ,\varphi \right) \right\rangle & =&\mathbf{R}
_{\theta ,\varphi }\left| j,-j\right\rangle =\frac{1}{\left( 1+\zeta ^{\ast
}\zeta \right) ^{j}}e^{\zeta \mathbf{J}_{+}}\left| j,-j\right\rangle \text{,}
\nonumber \\
& =&\sum_{m=-j}^{j}\frac{\zeta ^{j+m}}{\left( 1+\zeta ^{\ast }\zeta \right)
^{j}}\sqrt{\frac{\left( 2j\right) !}{\left( j+m\right) !\left( j-m\right) !}}
\left| j,m\right\rangle \text{,}  \label{estadocoerenteGeral}
\end{eqnarray}
where $\zeta =-e^{-i\varphi }\tan \frac{\theta }{2}$, with the angles $%
\varphi $\ and $\theta $\ as shown in Fig. 1, while $\mathbf{J}_{\pm }=%
\mathbf{J}_{x}\pm i\mathbf{J}_{y}$, and $\left| j,m\right\rangle $ are
eigenstates of the operator $\mathbf{J}_{z}$, with eigenvalues $m\hbar $.
The complex phase that characterizes the ACS may be defined in the complex
plane $Z$ (Greek capital $\zeta $) or over the Bloch sphere, where the
ground state is represented by the south pole and the most excited state is
identified with the north pole.

\begin{figure}[ptb]
\includegraphics[width=0.48\textwidth]{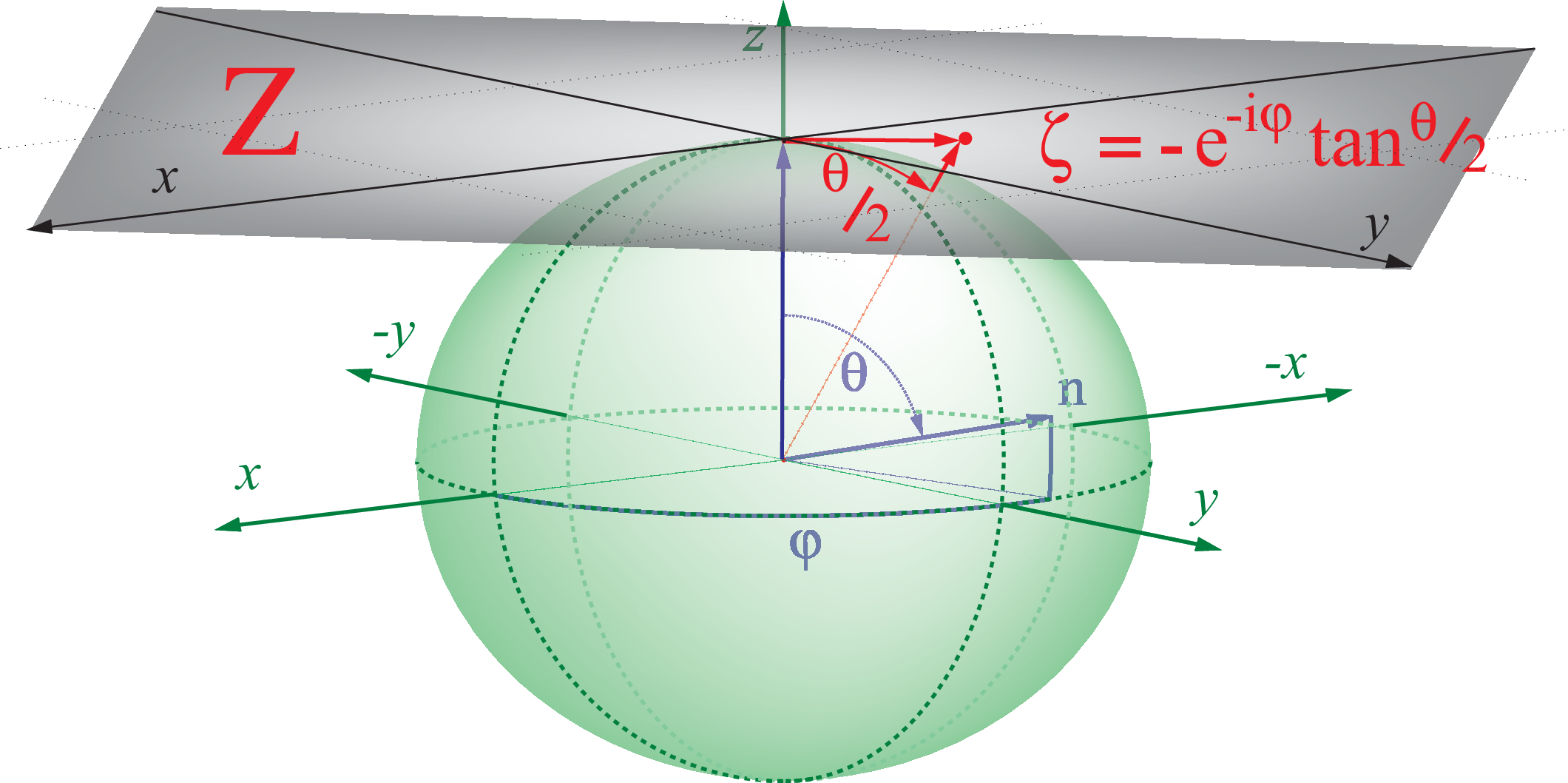}
\caption{(Color online) For a nuclear spin system we define rotation angles $\protect
\theta $ and $\protect\varphi $ on the Bloch sphere relative to the $\mathbf{
xyz}$ frame and \ the corresponding value of $\protect\zeta $ in the complex
plane denoted by $Z$, the Greek capital form of $\protect\zeta $. The north
pole corresponds to $\protect\zeta =0$ and the south pole to $\protect\zeta 
=\infty $.}
\label{fig:GeometriaEstadoCoerenteAtomico}
\end{figure}

Let us consider the density operator of the ACS denoted by $\rho =\left\vert
\zeta \left( \theta ,\varphi \right) \right\rangle \left\langle \zeta \left(
\theta ,\varphi \right) \right\vert $ $=\sum_{m,m^{\prime }=-j}^{j}\rho
_{m,m^{\prime }}\left\vert j,m\right\rangle \left\langle j,m^{\prime
}\right\vert $. From Eq. (\ref{estadocoerenteGeral}), each element of $\rho $
is given by 

\begin{eqnarray}
\rho_{m,m^{\prime}} & =&(-1)^{2j+m+m^{\prime}}e^{i(m^{\prime}-m)\varphi}
\cos^{2j-m-m^{\prime}}\left(  \theta / 2\right)   \sin^{2j+m+m^{\prime}}\left(
\theta / 2\right) \nonumber \\
&& \times\sqrt{\frac{\left(  2j\right)  !}{\left(  j+m^{\prime}\right)
!\left(  j-m^{\prime}\right)  !}} \sqrt{\frac{\left(  2j\right)  !}{\left(
j+m\right)  !\left(  j-m\right)  !}}\text{.} \label{ElementodeRhomn}
\end{eqnarray}

Hence, if the angular parameters $\theta $ and $\varphi $ are known, the
amplitude of each element $\rho _{m,m^{\prime }}$ can be obtained. The
inverse procedure, that is, the computation of $\theta $ and $\varphi $ from
the elements $\rho _{m,m^{\prime }}$, is also possible. In the case of $%
\theta $, we use

\begin{equation}
\begin{array}{ccc}
\sqrt[4j]{\rho_{j,j}} & = & \sin\frac{\theta}{2}\text{,} \\ 
\sqrt[4j]{\rho_{-j,-j}} & = & \cos\frac{\theta}{2}\text{,}
\end{array}
\label{phase01}
\end{equation}
for any value of $j$. In the case of $\varphi$, when $j$ is an integer

\begin{eqnarray}
\frac{\left( \rho_{0,1}+\rho_{1,0}\right) \sqrt{\left( j+1\right) !\left(
j-1\right) !}\left( j\right) !}{2\left( -1\right) ^{2j+1}\left( \cos\frac{%
\theta}{2}\right) ^{2j-1}\left( \sin\frac{\theta}{2}\right) ^{2j+1}\left(
2j\right) !} & =&\cos\varphi\text{,}  \label{phaseinteira02} \\
\frac{\left( \rho_{0,1}-\rho_{1,0}\right) \sqrt{\left( j+1\right) !\left(
j-1\right) !}\left( j\right) !}{2i\left( -1\right) ^{2j+1}\left( \cos\frac{%
\theta}{2}\right) ^{2j-1}\left( \sin\frac{\theta}{2}\right) ^{2j+1}\left(
2j\right) !} & =&\sin\varphi\text{.}  \label{phaseinteira05}
\end{eqnarray}
Similar expressions can be obtained in terms of the density matrix elements $%
\rho_{0,-1}$ and $\rho_{-1,0}$ (not shown). For $j$ a half integer, these
equations become

\begin{eqnarray}
\frac{\left( \rho_{\frac{-1}{2},\frac{1}{2}}+\rho_{\frac{1}{2},\frac{-1}{2}%
}\right) \left( j-\frac{1}{2}\right) !\left( j+\frac{1}{2}\right) !}{2\left(
-1\right) ^{2j}\left( \cos\frac{\theta}{2}\right) ^{2j}\left( \sin\frac{%
\theta}{2}\right) ^{2j}\left( 2j\right) !} & =&\cos \varphi\text{,}
\label{phaseinteirahalf02} \\
\frac{\left( \rho_{\frac{-1}{2},\frac{1}{2}}-\rho_{\frac{1}{2},\frac{-1}{2}%
}\right) \left( j-\frac{1}{2}\right) !\left( j+\frac{1}{2}\right) !}{%
2i\left( -1\right) ^{2j}\left( \cos\frac{\theta}{2}\right) ^{2j}\left( \sin%
\frac{\theta}{2}\right) ^{2j}\left( 2j\right) !} & =&\sin \varphi\text{.}
\label{phaseinteirahalf03}
\end{eqnarray}

Therefore a general ACS given by (\ref{estadocoerenteGeral}) is completely
characterized by $j$, $\theta$, and $\varphi$.

\section{NMR quadrupolar systems \label{sec:NMRQuadrupolar}}

NMR quadrupolar systems are composed of nuclei with spin $I>1/2$ subjected
to both a magnetic field and an electric field gradient, where the nuclear
magnetic moment is quantized as $M=I,I-1,...,-I$. Considering also the
interaction with a radio frequency (\textit{RF}) field used for excitation,
in a reference frame rotating around $z$-axis with angular velocity $\omega
_{RF}$, the NMR Hamiltonian can be written {(for more details see section
VII of Ref. \cite{abragam1994})}: 
\begin{eqnarray}
\mathcal{H}_{NMR}& =&-\hbar \left( \omega _{L}-\omega _{RF}\right) \mathbf{I}%
_{z}+\hbar \frac{\omega _{Q}}{6}\left( 3\mathbf{I}_{z}^{2}-\mathbf{I}%
^{2}\right)   \nonumber \\
&& +\hbar \omega _{1}\left( \mathbf{I}_{x}\cos \varphi _{s}+\mathbf{I}%
_{y}\sin \varphi _{s}\right) +\mathcal{H}_{env}\text{,}
\label{hamiltonianoRMNQ}
\end{eqnarray}
where $\mathbf{I}_{x}$, $\mathbf{I}_{y}$, $\mathbf{I}_{z}$ are the $x,y,z$
components of the total spin angular momentum operator. The first term of
the Hamiltonian is due to the interaction of the nuclear magnetic moments
with the strong magnetic field $B_{0}$ (Zeeman term). The second is due to
the interaction of the quadrupole moments of the nuclei with the internal
electric field gradient. $\left( \omega _{L}\right) $ and $\left( \omega
_{Q}\right) $ stand for the Larmor and Quadrupolar frequencies (coupling
strengths), which in this case, satisfy the inequality $\left| \omega
_{Q}\right| \ll \left| \omega _{L}\right| $. The third term represents the
external time-dependent \textit{RF} field perturbation of intensity $%
B_{1}=\omega _{1}/\gamma $, applied along the direction of the corresponding
vector $\mathbf{m}$, with $\gamma $ being the gyromagnetic ratio. The fourth
term represents weak interactions between quadrupolar nuclei and other
nuclei, electrons, random fields, referred to here as environment. This term
will be neglected as its contribution is weak \cite{oliveira2007}.

A quadrupolar system in the rotating frame will be used to describe two
different \ situations. In \ the \ first, the  external  time-dependent  perturbation
is \ present, with $\left| \omega _{L}\right| \cong \left| \omega _{RF}\right| 
$, and $\left| \omega _{1}\right| >>\left| \omega _{Q}\right| $ specifies a
regime where the perturbation equally affects all energy levels of the
system, non-selective pulse; the evolution operator is then given by

\begin{equation}
\mathcal{U}_{RF}\cong \exp \left[ -i\omega _{1}t\left( \mathbf{I}_{x}\cos
\varphi _{s}+\mathbf{I}_{y}\sin \varphi _{s}\right) \right] \text{.}
\label{RFPulse}
\end{equation}
The physical parameters to be identified are the polar $\left( \theta
\right) $ and azimuthal $\left( \varphi \right) $\ angles defining the
rotation operator $\mathbf{R}_{\theta ,\varphi }$. From the \textit{RF}
operator $\mathcal{U}_{RF}\left( t\right) $, these angles are 
\begin{eqnarray}
\theta & \equiv & \omega _{1}t\text{,}  \label{thetamudavartheta} \\
\varphi & \equiv & \varphi _{s}+\frac{\pi }{2}\text{.}  \label{phimudavarphi}
\end{eqnarray}

Secondly, when the external \textit{RF} perturbation is absent, the
corresponding evolution operator is 
\begin{equation}
\mathcal{U}_{Ev}=\exp \left[ i\left( \omega _{L}-\omega _{RF}\right) t%
\mathbf{I}_{z}-\frac{i\omega _{Q}t}{6}\left( 3\mathbf{I}_{z}^{2}-\mathbf{I}%
^{2}\right) \right] \text{,}  \label{FreeEvolution}
\end{equation}
which refers to a free evolution regime. Note that the operator $\mathcal{U}%
_{Ev}$ cannot be written as a rotation operator $\mathbf{R}_{\theta ,\varphi
}$, since it is not generated by the vectors $\mathbf{n}$ and $\mathbf{m}$
(see section \ref{sec:TSCSsection}). This approach is similar to that
discussed by Kitagawa and Ueda \cite{kitagawa1993} in the study of spin
squeezed states.

\section{Experimental implementation of the pseudo-NSCS \label%
{sec:experimentalprocesso}}

The NMR experiments were carried out at room temperature in a VARIAN INOVA
400 MHz spectrometer. The pseudo-NSCS was implemented with the $^{23}$Na
nuclei ($I=3/2$) present in sodium dodecil sulfate (SDS) in a lyotropic
liquid crystal, prepared with $21.3$ wt \% SDS, $3.7$ wt \% decanol, and $75$
wt \% D$_{2}$O. The strength of the Larmor frequency and quadrupolar
couplings were 105.85 MHz and 15 kHz, respectively. Typical $\pi $-pulse
lengths of 8 $\mu $s and recycle delays of 500 ms were used. The T$_{2}$ and
T$_{1}$ relaxation times of the $^{23}$Na nuclear spins are $2.6\pm 0.3$ ms
and $12.2\pm 0.2$ ms, respectively.

\begin{figure}[tbp]
\includegraphics[width=0.5\textwidth]{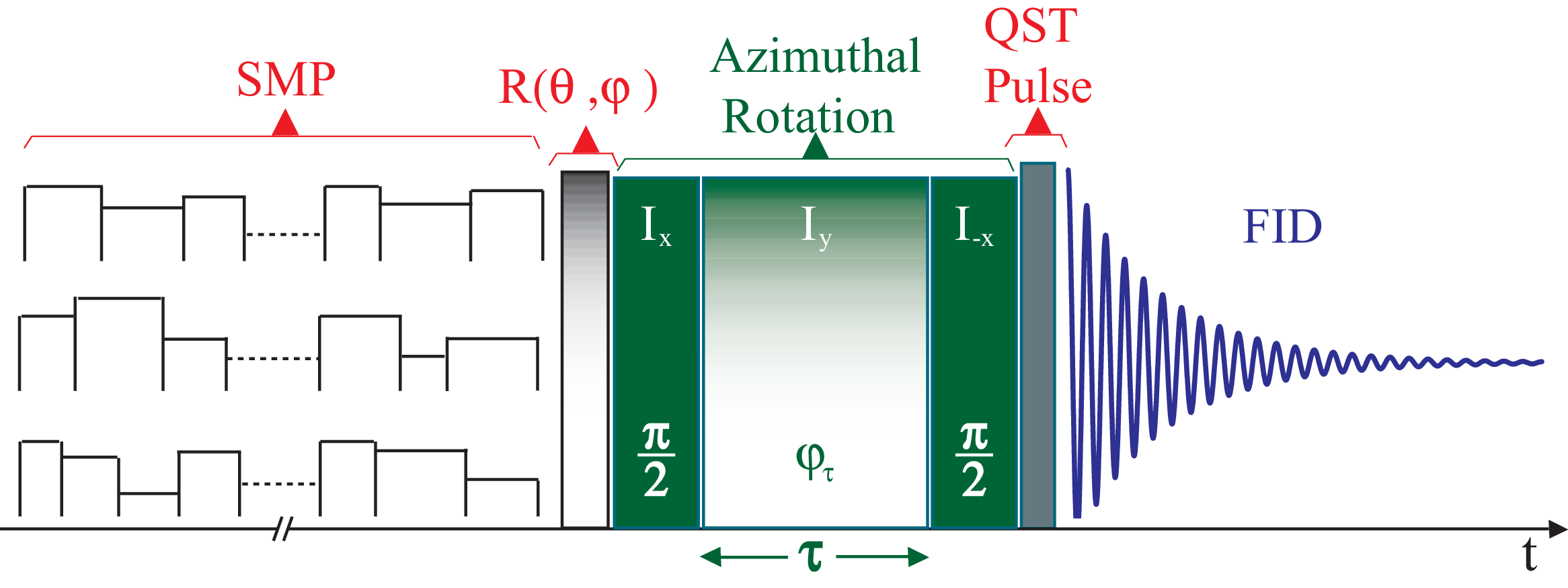}
\caption{(Color online) The pulse sequence has four steps. In the first step the
pseudo-NSCS $\left\vert I,I\right\rangle $ is implemented by the SMP
technique. The second step consists of the non-selective pulses with
parameters $\left( \protect\theta ,\protect\varphi \right) $ that produce
excited pseudo-NSCSs $\left\vert \protect\zeta \left( \protect\theta ,%
\protect\varphi \right) \right\rangle $. The third step is the azimuthal
rotation through time $\protect\tau $, which can be used to produce cyclic
evolutions. The last step involves the read-out of the pseudo-NSCS (QST
procedure).}
\label{fig:SequenciaPulsos}
\end{figure}

Room temperature thermal equilibrium NMR systems are almost maximum mixture
states, represented by a density matrix which deviates only slightly from
the normalized identity matrix 
\begin{equation}
\rho \approx \frac{\left( \mathbf{1}-\beta {\hbar \omega _{L}}\mathbf{\ I}%
_{z}\right) }{{\mathcal{Z}}}\text{,}  \label{densitymatrix}
\end{equation}
where $\beta =1/k_{B}T$ and ${\mathcal{Z}}={\mathtt{Tr}\left[ e^{\left(
-\beta \mathcal{H}_{NMR}\right) }\right] }$ is the partition function, $T$
the room temperature and $k_{B}$ the Boltzmann constant. {Using suitable
spin rotations -- \textit{RF} pulses and free evolutions -- and a temporal
average technique (so-called strong modulated pulses - SMP \cite
{fortunato2002})}, this thermal state can be transformed into a state of the
form

\begin{equation}
\rho \approx \left( \frac{1}{\mathcal{Z}}-\epsilon \right) \mathbf{1}%
+\epsilon \left\vert \psi \right\rangle \left\langle \psi \right\vert ,
\label{pseudopurestate}
\end{equation}
where $\left\vert \psi \right\rangle \left\langle \psi \right\vert $ is
called a pseudo-pure state, having the form of a pure state density matrix
with unitary trace, and $\epsilon =\beta {\hbar \omega _{L}/}\mathcal{Z}$.
By the technique of quantum state tomography (QST) \cite{teles2007}, we
obtain experimentally the quantum state $\left\vert \psi \right\rangle $
that will be used to represent the pseudo-NSCS $\left\vert I,I\right\rangle
\equiv \left\vert \zeta \left( 0,0\right) \right\rangle $. This fundamental
pseudo-pure state represents, from the NMR point of view, the precession of
the magnetic moment around an axis defined by the direction of the strong
static magnetic field $B_{0}$. To obtain other excited pseudo-NSCSs, we just
need to apply the operator $\mathcal{U}_{RF}$ (Eq. (\ref{RFPulse})) to $%
\left\vert I,I\right\rangle $. In Fig. \ref{fig:SequenciaPulsos} we show how
to implement the fundamental and excited pseudo-NSCSs.

\subsection{Initializing pseudo-NSCSs}

To produce excited pseudo-NSCSs from $\left\vert I,I\right\rangle $, we
apply non-selective pulses of strength $\omega _{1}$ and phase $\varphi _{s}$
during time interval $t=\theta /\omega _{1}$ (see Eq. (\ref{RFPulse})). As a
first example, the excited pseudo-NSCS $\left\vert \zeta \left( \pi
/2,0\right) \right\rangle =\left\vert -1\right\rangle $ can be generated by
applying a non-selective $\pi /2$ pulse that produces a rotation about the
negative $y$--axis. On the other hand, a rotation of $\pi /2$ about the
positive $y$--axis of the vector state $\left\vert I,I\right\rangle $
generates the state $\left\vert \zeta \left( \pi /2,\pi \right)
\right\rangle =\left\vert 1\right\rangle $. We observe that the unitary
transformation (of Eq. (\ref{RFPulse})) associated with non-selective
pulses, acting on a given pseudo-NSCS, produces excited pseudo-NSCSs
analogously to those discussed in Ref. \cite{duzzioni2007}.

In Fig. 3, we present experimental results for the deviation density matrix,
in the form of bar charts of the real (left) and imaginary (middle)
components, for the quantum states $\left| \zeta \left( 0,\varphi \right)
\right\rangle =\left| 0\right\rangle $, $\left| \zeta \left( \pi /2,\pi
\right) \right\rangle =\left| 1\right\rangle $, $\left| \zeta \left( \pi
/2,0\right) \right\rangle =\left| -1\right\rangle $, \  $\left| \zeta \left(
\pi /2,3\pi /2\right) \right\rangle =\left| i\right\rangle $, \ $\left| \zeta
\left( \pi /2,\pi /2\right) \right\rangle =\left| -i\right\rangle $, and $%
\left| \zeta \left( \pi ,\varphi \right) \right\rangle =\left| \infty
\right\rangle $ (see Tab. \ref{tab:angularvalues}). The intensity of the
elements of each deviation density matrix was obtained by QST. On the right
of Fig. 3, the Husimi $\mathcal{Q}$-distribution was used to characterize
these implemented quantum states, which is sketched on the Bloch sphere \cite
{gordon1999}. The highest positive normalized intensity corresponds to the
red color and the most negative intensity corresponds to the orange color.
The highest intensity (red color) of the quasi-probability distribution
function indicates the orientation of the spin nuclei magnetization. The
large dispersion of the Husimi $\mathcal{Q}$-distribution function is due to
the small spin value $I=3/2$, or analogously, the small number of particles $%
N=2I=3$ of an atomic system.

\begin{figure}[ptb]
\caption{(Color online) Experimental results for the $^{23}$Na nuclei in the
sample of SDS. The QST results are represented by bar charts (left and
middle column) for the implemented pseudo-NSCSs and on the Bloch Sphere
(right column) by the Husimi $\mathcal{Q}$-distribution. The top bar chart
represents the pseudo-NSCS $\left| 0\right\rangle $, the left (right) bar
chart showing the real (imaginary) part of the deviation density operator,
and similarly for the other bar charts. The pseudo-NSCS $\left|
0\right\rangle $ is implemented as explained in section \ref
{sec:experimentalprocesso} by the SMP technique. The bottom bar chart shows
the pseudo-NSCS $\left| \infty \right\rangle $, a state that is implemented
by applying a rotation operator $\mathbf{R}_{\protect\theta ,\protect\varphi
}$ with $\protect\theta =\protect\pi $ and $\protect\varphi =3 \protect\pi %
/2 $ to the pseudo-NSCS $\left| 0\right\rangle $. The pseudo-NSCS $\left|
-1\right\rangle $ is implemented by applying a ${\protect\pi }/{2}$
non-selective pulse in the negative $y-$direction to transform $\left|
0\right\rangle $. Analogous remarks apply to the bar charts labeled as $%
\left| 1\right\rangle $, $\left| i\right\rangle $, and $\left|
-i\right\rangle $. In the right column, we present the calculated Husimi $%
\mathcal{Q}$-distribution of each deviation density matrix and, at the
bottom, the intensity of the $\mathcal{Q}$--distribution function is encoded
in the colour bar.}
\label{fig:GeometriaEstadoCoerenteSpinValor}\includegraphics[width=0.45		%
\textwidth]{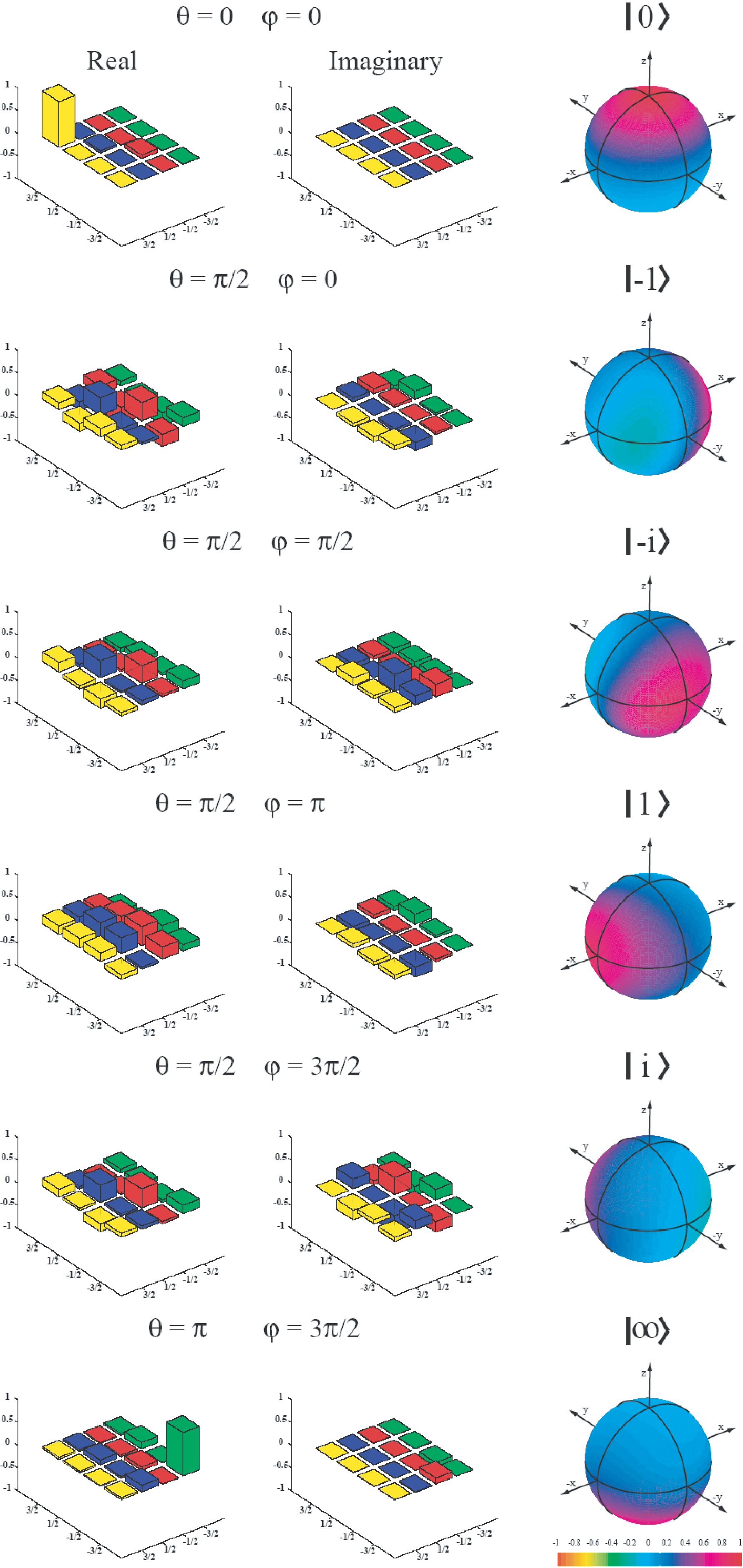}
\end{figure}

In order to compare the theoretical predictions with the experimental values
of the deviation density matrices of Fig. 3, we present in Tab. \ref
{tab:angularvalues} the complex phase $\zeta \left( \theta ,\varphi \right) $
and the Bloch vector $\mathbf{n}$. The experimental values for $\theta $ and 
$\varphi $ were computed with Eqs. (\ref{phase01}), (\ref{phaseinteirahalf02}%
), and (\ref{phaseinteirahalf03}).

\begin{table*}[hbp]
\centering
\label{tab:angularvalues}
\begin{tabular}{c|c|c|c|c|c}
\hline\hline
\multicolumn{2}{l|}{} & \multicolumn{2}{|c|}{Theoretical} & 
\multicolumn{2}{|c}{Experimental} \\ \hline
$\theta$ & $\varphi$ & $\left| \zeta\left( \theta,\varphi\right)
\right\rangle $ & $\mathbf{n}$ & $\left| \zeta\left( \theta,\varphi\right)
\right\rangle $ & $\mathbf{n}$ \\ \hline
$0$ & $0$ & $\left| 0\right\rangle $ & $\left( 0,0,1\right) $ & $\left|
0.02-0.01i\right\rangle $ & $\left( -0.05,-0.02,0.99\right) $ \\ 
$\frac{\pi}{2}$ & $0$ & $\left| -1\right\rangle $ & $\left( 1,0,0\right) $ & 
$\left| -0.88-0.12i\right\rangle $ & $\left( 0.98,-0.14,0.12\right) $ \\ 
$\frac{\pi}{2}$ & $\frac{3\pi}{2}$ & $\left| -i\right\rangle $ & $\left(
0,-1,0\right) $ & $\left| -0.88i\right\rangle $ & $\left(
0,-0.99,0.13\right) $ \\ 
$\frac{\pi}{2}$ & $\pi$ & $\left| 1\right\rangle $ & $\left( -1,0,0\right) $
& $\left| 0.79+0.25i\right\rangle $ & $\left( -0.94,0.30,0.18\right) $ \\ 
$\frac{\pi}{2}$ & $\frac{\pi}{2}$ & $\left| i\right\rangle $ & $\left(
0,1,0\right) $ & $\left| -0.09+0.84i\right\rangle $ & $\left(
0.1,0.98,0.16\right) $ \\ 
$\pi$ & $0$ & $\left| \infty\right\rangle $ & $\left( 0,0,-1\right) $ & $%
\left| 7.75+1.16i\right\rangle $ & $\left( -0.25,0.04,-0.97\right) $ \\ 
\hline\hline
\end{tabular}
\vspace*{0.5cm}  
\caption{Comparison between theoretical and experimental values obtained
from the deviation density matrix of Fig. 3 by means of equations (\ref
{phase01}), (\ref{phaseinteirahalf02}) and (\ref{phaseinteirahalf03}).
Experimental values are rounded to two decimal places. }
\end{table*}

\subsection{Controlling pseudo-NSCSs}

Besides knowing how to initialize a quantum system, i.e., how to prepare the
desired quantum state, an important task in quantum physics is the control
of the quantum system with high accuracy. The implementation and control of
pseudo-NSCSs are important for applications in quantum computing and
geometric phases \cite{duzzioni2007}, squeezed states \cite{kitagawa1993},
and quantum metrology \cite{riedel2010}. In the present case, such control
is performed through polar and azimuthal rotations.

\textit{Polar rotations} -- In the context of polar rotations, we have
demonstrated the experimental control of angle $\theta $ by taking the
initial quantum state $\left\vert \zeta \left( 0,0\right) \right\rangle
=\left\vert 0\right\rangle $ and applying the rotation operator $\mathbf{R}%
_{\theta ,\varphi }$, with \ $\theta =0,\pi /18,2\pi /18,$ $\ldots ,\pi $
(total of nineteen steps) and fixed $\varphi =\pi $. After each step, the
state $\left\vert \zeta \left( \theta ,\varphi \right) \right\rangle $ was
reconstructed by QST (see pulse sequence in Fig. 2). First, the average
values of the magnetic angular momenta $\left\langle \mathbf{I}%
_{x,y,z}\right\rangle $\ were calculated for each deviation density matrix $%
\left\vert \zeta \left( \theta ,\varphi \right) \right\rangle \left\langle
\zeta \left( \theta ,\varphi \right) \right\vert $. The experimental results
(symbols) are shown in Fig. 4 together with the values obtained by numerical
simulation (solid lines) and theoretical prediction (dashed lines). We use
the term numerical simulation when we apply numerical recipes to the
dynamics of evolution given by Hamiltonian (\ref{hamiltonianoRMNQ}), while
the theoretical prediction is given by the evolution operator of Eq. (\ref
{RFPulse}). The differences between simulated data and theoretical
prediction are caused by the quadratic term of Hamiltonian (\ref
{hamiltonianoRMNQ}). Such a term is the main source of imperfections in the
generation and control of the pseudo-NSCSs.

In order to find the fidelity of the experimental implementations, we
compared the tomographed deviation density matrix $\rho _{Exp}$ with the
theoretical prediction $\rho _{The}$, by calculating \cite{fortunato2002} 
\begin{equation}
F=\frac{{\mathtt{Tr}}\left\{ \rho _{Exp}\cdot \rho _{The}\right\} }{\sqrt{{%
\mathtt{Tr}}\left\{ \rho _{Exp}\cdot \rho _{Exp}\right\} {\mathtt{Tr}}%
\left\{ \rho _{The}\cdot \rho _{The}\right\} }}\text{.}
\label{CorrelacionExpTheo}
\end{equation}
The results for $F$ are presented in Fig. 4b and show that the experimental
generation and control of the pseudo-NSCSs can be achieved with fidelity
higher than 80 \%.

\begin{figure}[tbp]
\includegraphics[width=0.5\textwidth]{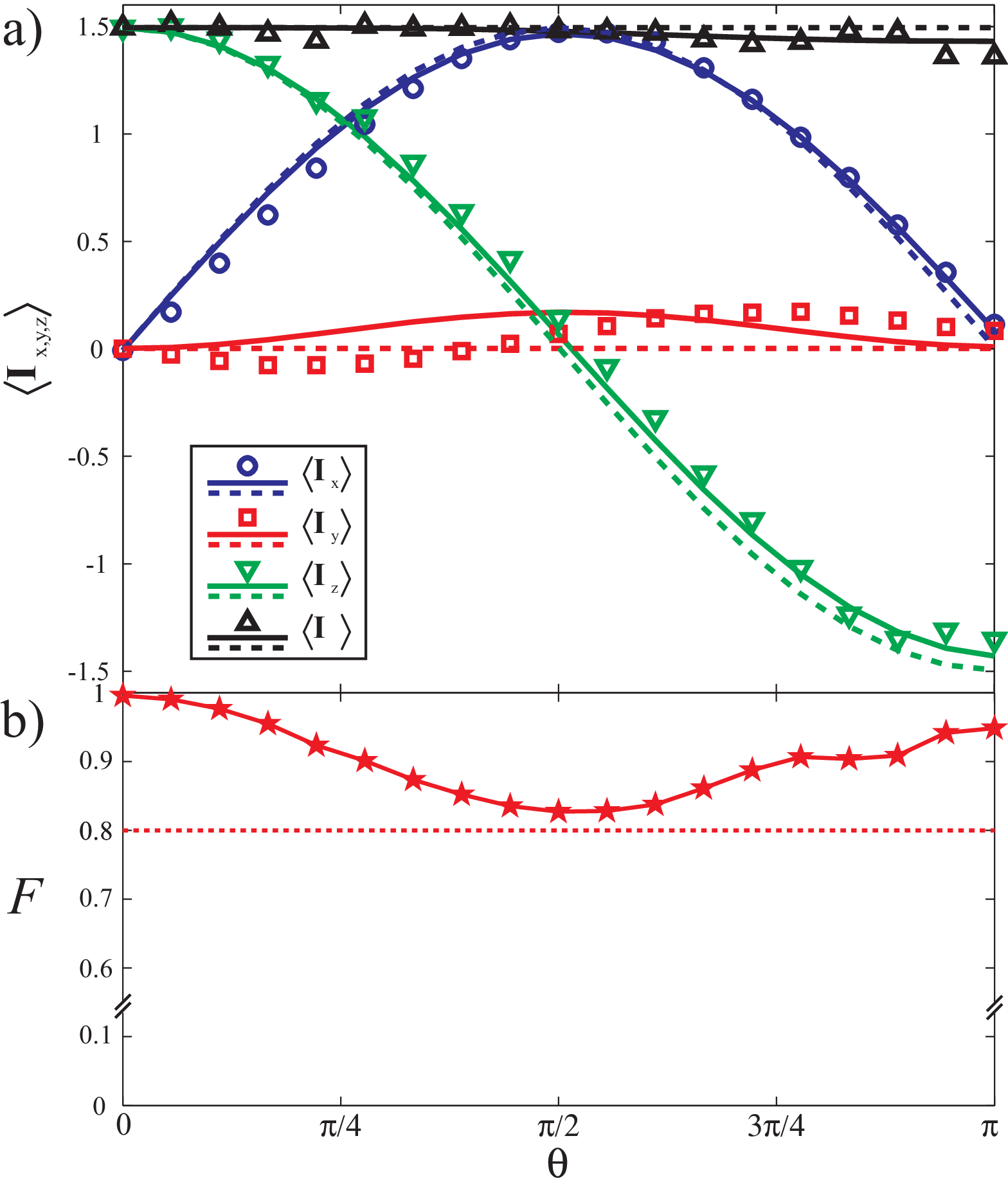}
\caption{(Color online) Experimental results (symbols), numerical simulations (solid
line), and theoretical predictions (dashed lines) of a polar rotation of the
Bloch vector for the initial pseudo-NSCS $\left| 0\right\rangle $,
implemented experimentally by the SMP technique. QST is performed for
nineteen different values of $\protect\theta $ from 0 to $\protect\pi $. The
average values of the magnetic angular momentum components $\left\langle 
\mathbf{I}_{x,y,z}\right\rangle $ and the total angular momentum $%
\left\langle \mathbf{I}\right\rangle =\protect\sqrt{\sum_{i=x,y,z}\left%
\langle \mathbf{I}_{i}\right\rangle ^{2}}$ are shown in a). The accuracy of
the polar rotations is calculated by the fidelity $F$ (Eq. (\ref
{CorrelacionExpTheo})) between experimental results and theoretical
prediction, as depicted in b).}
\label{fig:PolarValorMedio}
\end{figure}

\textit{Azimuthal rotations.--} The azimuthal rotation is performed around
the $z$-axis so that it\ can be achieved by the pulse sequence $\left[ \pi /2%
\right] _{x}\rightarrow \left[ \varphi _{\tau }\right] _{y}\rightarrow \left[
\pi /2\right] _{-x}$, where the subscripts represent the phases of the
non-selective pulses and the values in square brackets the rotation angles.
This pulse sequence is referred to as composite pulse \cite{freeman1981}. As
a first step in the experiment, we arbitrarily prepare the excited
pseudo-NSCS $\left| \zeta \left( \pi /2,\pi \right) \right\rangle $. By
applying the composite pulse sequence to the initial state $\left| \zeta
\left( \pi /2,\pi \right) \right\rangle $ we can obtain thirty-three
distinct pseudo-NSCSs $\left| \zeta \left( \pi /2,\pi +\varphi _{\tau
}\right) \right\rangle $ with angular values $\varphi _{\tau }=0,\pi
/16,2\pi /16,3\pi /16,\ldots ,2\pi $ and $\theta $ fixed at $\pi /2$.
Finally, the transformed state is reconstructed by the QST procedure.

The average values of magnetic angular momenta $\left\langle \mathbf{I}%
_{x,y,z}\right\rangle $ \ were \ computed \ for \ each \ deviation \ density \ matrix $%
\left| \zeta \left( \pi /2,\pi +\varphi _{\tau }\right) \right\rangle
\left\langle \zeta \left( \pi /2,\pi +\varphi _{\tau }\right) \right| $. The
experimental results (symbols) are plotted in Fig. 5a together with the
values of the numerical simulation (solid lines) and theoretical prediction
(dashed lines). As in the case of polar rotation, using one non--selective
pulse, the differences between simulated data and theoretical prediction are
caused by the quadratic term of Hamiltonian (\ref{hamiltonianoRMNQ}).
However, in the study of azimuthal rotation we need four pulses: one to
create the excited state $\left| \zeta \left( \pi /2,\pi \right)
\right\rangle $ and three making up the composite pulse sequence. These
additional pulses are another source of errors in the generation of the
state, due to the error accumulated during the calibration of each pulse and
time delays between them, which are of order 2--3 $\mu $s, despite being
small, is not truly negligible. In the azimuthal rotation, the extra pulses
and delays take 10--12 $\mu $s more time to implement than a polar rotation
by the same angle. During this additional time, the contribution of the
quadrupolar coupling is more evident, transforming the implemented
pseudo-NSCS with a non-null degree of squeeze \cite{kitagawa1993}. This is
the reason why the fidelity $F$ in Fig. 5b is, on average, lower than the
fidelity in the polar case (see Fig. 4b). In order to diminish these errors
it is necessary to calibrate non--selective pulses of short duration.

\begin{figure}[ptb]
\includegraphics[width=0.48\textwidth]{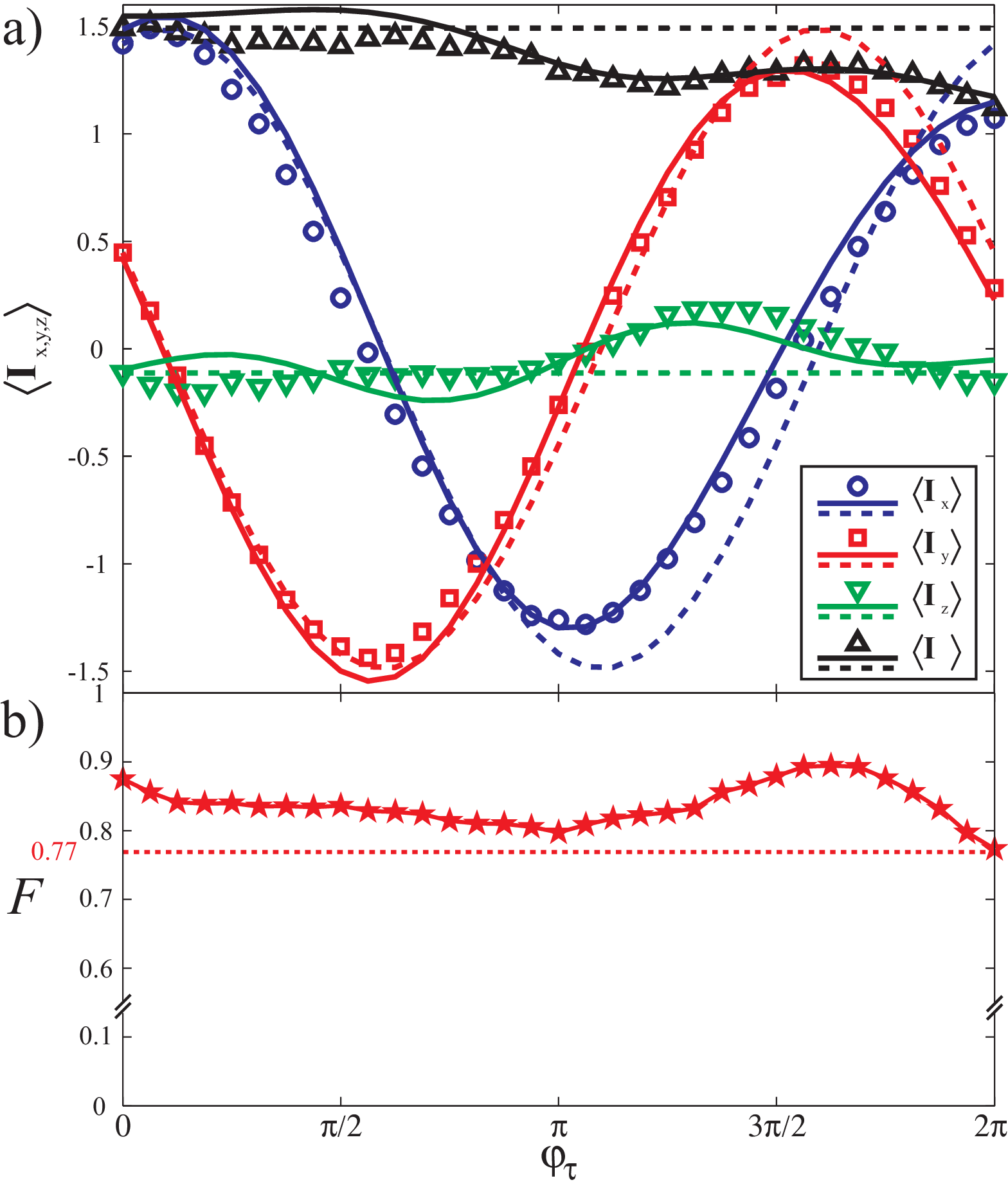}
\caption{(Color online) Experimental results (symbols), numerical simulations (solid
line), and theoretical predictions (dashed lines) of an azimuthal rotation
of the Bloch vector for the initial pseudo-NSCS $\left| 1\right\rangle $,
implemented experimentally by the SMP technique. QST is performed for
thirty-three different values of $\protect\varphi _{\protect\tau }$ from 0
to $2\protect\pi $. The average values of the magnetic angular momentum
components $\left\langle \mathbf{I}_{x,y,z}\right\rangle $ and the total
angular momentum $\left\langle \mathbf{I}\right\rangle =\protect\sqrt{%
\sum_{i=x,y,z}\left\langle \mathbf{I}_{i}\right\rangle ^{2}}$ are shown in
a). The accuracy of the azimuthal rotations is calculated by the fidelity $F$
(Eq. (\ref{CorrelacionExpTheo})) between experimental results and
theoretical prediction as depicted in b).}
\label{fig:AzimutalValorMedio}
\end{figure}

\section{Theoretical application of the pseudo-NSCS \label{sec:aplicaciones}}

\subsection{Geometric phase \label{sec:FaseGeometrica}}

In this section, we use the Mukunda-Simon formalism \cite{mukunda1993} to
compute geometric phases that arise from the three distinct cyclic
evolutions. For a system evolving according to the Schr\"{o}dinger equation,
the geometric phase $\phi _{G}$ can be defined as the difference between the
total phase $\phi _{T}\left( \tau \right) =\arg \left[ \left\langle \psi
\left( 0\right) |\psi \left( \tau \right) \right\rangle \right] $ and \ the \
dynamic \ phase $\phi _{D}\left( \tau \right) =-i\int_{0}^{\tau }\left\langle
\psi \left( t\right) \right\vert \frac{\partial }{\partial t}\left\vert \psi
\left( t\right) \right\rangle $, i.e., 
\begin{equation}
\phi _{G}\left( \tau \right) =\arg \left[ \left\langle \psi \left( 0\right)
|\psi \left( \tau \right) \right\rangle \right] +i\int_{0}^{\tau
}\left\langle \psi \left( t\right) \right\vert \frac{\partial }{\partial t}%
\left\vert \psi \left( t\right) \right\rangle dt\text{,}
\label{PhaseGeometricaMukunda}
\end{equation}
where $\tau $ is the overall time of the evolution \cite
{duzzioni2007,duzzioni2005,fuentes2002}.

Before starting the calculation of the geometric phases, it is worth
mentioning that this phase, acquired by the quadrupolar NMR system, is
related only to the pseudo-pure state, since its definition (\ref
{PhaseGeometricaMukunda}) does not hold for general mixed states \cite
{tong2004PRL}. Assuming an initial pseudo-NSCS $\left\vert \zeta \left(
\theta ,\varphi \right) \right\rangle $, the geometric phase acquired by
this quantum state is calculated while undergoing three distinct cyclic
evolutions: $i)$ the free evolution under the NMR quadrupolar Hamiltonian
(Eq. (\ref{hamiltonianoRMNQ})), and the evolution of such a system under the
analogous $ii)$ single- and $iii)$ two-mode BEC Hamiltonian. This analogy
between the NMR qua\-dru\-po\-lar system \ and \ BECs \ was \ explored \ in Ref. \cite
{auccaise2009} through the Holstein-Primakov formalism.

\subsection{Free evolution of the NMR system \label{sec:FreeEvolucionNMR}}

The first case to be analyzed is the free evolution of the quadrupolar
system with the offset frequency set $\omega _{RF}=\omega _{L}-\omega _{Q}$.
In this case the evolution operator, Eq. (\ref{FreeEvolution}), is rewritten
as 
\begin{equation}
\mathcal{U}_{Ev}\left( \varphi ^{\prime }\right) =\exp \left[ -i\varphi
^{\prime }\left( -\mathbf{I}_{z}+\frac{1}{6}\left( 3\mathbf{I}_{z}^{2}-%
\mathbf{I}^{2}\right) \right) \right] \text{,}  \label{NMRQuadrupolarUev}
\end{equation}
where $\varphi ^{\prime }=\omega _{Q}t$, which from now on will vary from $0$
to $2\pi $. Using Eq. (\ref{PhaseGeometricaMukunda}) and computing the
average values of operators $\mathbf{I}_{z}$ and $\mathbf{I}$ for the
pseudo-NSCS $\left\vert \zeta \left( \theta ,\varphi \right) \right\rangle $%
, the geometric phase can be expressed by: 
\begin{equation}
^{A}\phi _{G}\left( \mathcal{C}\right) =\pi I\left[ \left( \left( I-\frac{1}{%
2}\right) \cos ^{2}\theta -2\cos \theta +\frac{5}{2}-I\right) \right] \text{.%
}  \label{FasGeoA}
\end{equation}
This expression has a quadratic dependence on the harmonic function, because
the Hamiltonian in this analysis is quadratic in the nuclear spin angular
momentum operator. Applying Eq. (\ref{FasGeoA}) to a spin $I=1/2$ nucleus,
we find 
\begin{equation}
^{A}\phi _{G}\left( \mathcal{C}\right) =\pi \left( 1-\cos \theta \right) 
\text{,}  \label{FaseGeometricaSpinQuadrupolar}
\end{equation}
which corroborates the well known results reported for two-level particles
in references \cite{jones2000,ekert2000,leek2007,shapere}. Note that in this
case the single harmonic function arises from the linear dependence of the
nuclear spin angular momentum operator in the Hamiltonian, which allows the
connection between the geometric phase and half of the solid angle on the
Bloch sphere. On the other hand, for spin $I=3/2$, the presence of the
quadrupolar term makes the geometric phase proportional to the square of the
solid angle.

\subsection{NMR system evolving as a single-mode BEC}

As a second example of the free evolution of the NMR quadrupolar system, we
explore its analogy with the single-mode BEC, as reported in Ref. \cite
{auccaise2009}. In this description, the frequency $\omega _{RF}$ of the
evolution operator in Eq. (\ref{FreeEvolution}) satisfies the mathematical
expression $2I-1=\frac{2\left( \omega _{L}-\omega _{RF}\right) }{\omega _{Q}}
$ (see Eq. (7) of Ref. \cite{auccaise2009}). Then $\mathcal{U}_{Ev}\left(
\varphi ^{\prime }\right) $ is written: 
\begin{equation}
\mathcal{U}_{Ev}\left( \varphi ^{\prime }\right) =\exp \left[ -i\frac{%
\varphi ^{\prime }}{2}\left( \mathbf{I}_{z}^{2}-\left( 2I-1\right) \mathbf{I}%
_{z}-\frac{I^{2}+I}{3}\mathbf{1}\right) \right] \text{,}
\label{BECsingleUev}
\end{equation}
where the angular parameter is denoted by $\varphi ^{\prime }=\omega _{Q}t$.
The geometric phase acquired by the pseudo-NSCS $\left\vert \zeta \left(
\theta ,\varphi \right) \right\rangle $ transformed by the operator (\ref{BECsingleUev})\ is 
\begin{equation}
^{B}\phi _{G}\left( \mathcal{C}\right) =\frac{\pi I\left( 2I-1\right) }{2}%
\left( 3+\cos \theta \right) \left( 1-\cos \theta \right) \text{.}
\label{FasGeoB}
\end{equation}
In this interpretation, the geometric phase displays a qua-dra-tic dependence
on the harmonic function, which is linked to the quadratic dependence of the
number operator in the Holstein-Primakov representation \cite{mead1983}. We
note that for $I=1/2$ the geometric phase is null. This is consistent with a
discussion on geometric phase in a study of the quasiparticle dynamics in a
BEC where one particle becomes a free particle and is not part of the
condensate \cite{zhang2006PRL}.

\subsection{NMR system evolving as a two-mode BEC}

In the third case, the two-mode BEC Hamiltonian in the Schwinger
representation (see Eq. (14) of Ref. \cite{chen2004}) is simulated by
Hamiltonian (\ref{hamiltonianoRMNQ}) of the NMR system. In this \ description,
the \ pseudo-angular \ momentum \ operators $\left( \mathbf{J}_{i=x,y,z}\right) $
are mapped on to the nuclear spin momentum operators $\left( \mathbf{I}%
_{i=x,y,z}\right) $, and the Hamiltonian parameters are mapped by $\omega
_{0}\rightarrow \left( \omega _{RF}-\omega _{L}\right) $, $q\rightarrow
\omega _{Q}/2$, $G\rightarrow \omega _{1}$ and $\varphi \rightarrow \varphi
_{s}$. We will work in a particular configuration where the scattering
lengths satisfy the relation $A_{a}:A_{ab}:A_{b}=1.03:1:0.97$ for modes $a$
and $b$ of the BEC \cite{duzzioni2007}, implying $q=0$. This is usually
achieved through Feshbach resonances, as discussed in references \cite
{leggett2001,duzzioni2007,fuentes2002,chen2004}. We also impose the
condition that the external laser field coupling the two modes is turned off
($G=0$). The net atomic evolution is a rotation around the $z$--axis of
frequency $\omega _{0}$, which one in the NMR scenario is implemented by
means of the composite pulse sequence explained in section \ref
{sec:experimentalprocesso}, i.e. 
\begin{equation*}
\mathcal{U}\left( \varphi ^{\prime }\right) =\left[ \frac{\pi }{2}\right]
_{x}\longrightarrow \left[ \varphi ^{\prime }\right] _{y}\longrightarrow %
\left[ \frac{\pi }{2}\right] _{-x}\text{,}
\end{equation*}
where the pulse sequence is read from left to right. After a cyclic
evolution of state $\left| \zeta \left( \theta ,\varphi \right)
\right\rangle $, the geometric phase is: 
\begin{equation}
^{C}\phi _{G}\left( \mathcal{C}\right) =-2\pi I\left( 1-\cos \theta \right) 
\text{,}  \label{FasGeoC}
\end{equation}
reproducing a similar mathematical expression in Eq. (14) of Ref. \cite
{fuentes2002} for an adiabatic evolution. In this case the geometric phase
displays a linear dependence on the harmonic function, which reflects the
linear dependence of the Hamiltonian in the nuclear spin angular momentum
operator.

As in the Section \ref{sec:FreeEvolucionNMR}, the particular case of the
geometric phase for a nuclear spin $I=1/2$ gives: 
\begin{equation*}
^{C}\phi _{G}\left( \mathcal{C}\right) =-\pi \left( 1-\cos \theta \right) 
\text{,}
\end{equation*}
a result that resembles the previous expression (\ref{FaseGeometricaSpinQuadrupolar}) for a spin $I=1/2$ particle. The negative
sign means the opposite direction to the nuclear spin rotation. Indeed, this
is due to the equivalence between superposition of spin $\left\vert \uparrow
\right\rangle $ and spin $\left\vert \downarrow \right\rangle $ and the ACS
definition for a two level-system.

\section{Conclusion \label{sec:discusiones}}

In this paper we have described how it is possible to transfer the ACS
concepts to the nuclear quadrupolar system labeled as a pseudo-NSCS. We
experimentally showed how to initialize and to tomograph the pseudo-NSCS.
The experimental manipulation of this state was performed by polar and
azimuthal rotations, in which we note that the greatest obstacle to perfect
initialization and control is the quadrupolar term of the NMR Hamiltonian.
We also studied theoretically the geometric phase generated by the evolution
of the pseudo-NSCS in three cases: $i)$ free evolution of the NMR system, $%
ii)$ single and $iii)$\ two mode BEC. In the first and second cases, the
geometric phase is a quadratic function of $\cos \theta $ for $I>1/2$, which
is a signature of the quadrupolar term. For $I=1/2$, the geometric phase is $%
\pi (1-\cos \theta )$ for free evolution of the NMR system and null for the
single mode BEC case. In the latter case, for $\theta >0$, the null
geometric phase indicates there is no condensation. In the third case, the
geometric phase acquired by the pseudo-NSCS under the action of a composite
pulse sequence reproduces the same result as a geometric phase generated by
a cyclic azimuthal rotation. The present study would be still more complete
if one could measure experimentally the geometric phase discussed in each
case. This task could be undertaken with two coupled nuclear spin systems,
one spin $I>1/2$ and the other $S\geq 1/2$ (by analogy to Ref. \cite
{jones2000}), which we will try to implement in the near future. Finally,
the concept of pseudo--NSCS in the NMR quadrupolar scenario offers new
insights into spin squeezed states and quantum metrology implemented in
liquid crystals or  solid matter.

\begin{acknowledgments}
The authors acknowledge the financial support of the Brazilian Science Foundations CAPES, CNPq, FAPESP, FAPEMIG  and the Brazilian National Institute for Science and Technology of Quantum Information (INCT-IQ).
\end{acknowledgments}

\end{document}